\documentclass[twocolumn,showpacs,preprintnumbers,amsmath,amssymb]{revtex4}

\usepackage{graphicx}% Include figure files
\usepackage{dcolumn}% Align table columns on decimal point
\usepackage{bm}% bold math

\begin{document}

\title{A stochastical model for periodic 
domain structuring in ferroelectric crystals}

\author{Felix Kalkum}
  \email{kalkum@physik.uni-bonn.de}
\author{Helge A. Eggert}%
\author{Tobias Jungk}%
\author{Karsten Buse}
  \affiliation{Institute of Physics, University of Bonn, Wegelerstr.\ 8, 53115 Bonn, Germany.}

\date{\today}

\begin{abstract}
A stochastical description is applied in order 
to understand how ferroelectric structures can 
be formed. The predictions are compared with 
experimental data of the so-called electrical 
fixing: Domains are patterned in photorefractive 
lithium niobate crystals by the combination of 
light-induced space-charge fields with externally 
applied electrical fields. 
In terms of our stochastical model the probability for 
domain nucleation is modulated according to the sum
of external and internal  
fields. The model describes the shape of the domain 
pattern as well as the effective degree of modulation. 
\end{abstract}

\pacs{77.80,42.65.Hw}% PACS, the Physics and Astronomy
                             % Classification Scheme.
%\keywords{Suggested keywords}%Use showkeys class option if keyword
                              %display desired
\maketitle

% \section{Introduction}

The importance of ferroelectric optical materials 
is still contrasted by a lack of theoretical 
understanding. Comparing  textbook knowledge 
with real experimental results yields many 
discrepancies. Some examples for the prominent 
ferroelectric materials lithium niobate and 
lithium tantalate: Domain walls can be pinned 
and can bow \cite{Yang99Dire}, the coercive field 
can be reduced by an order of magnitude by small
changes in the lithium content \cite{Kitamura98Crys}, 
and the materials exhibit a memory effect for 
recent domain inversions \cite{Brown99Cont}. 
The reason that it is so hard to predict and to 
model ferroelectric domain inversion and patterning 
results from the influence of defects (vacancies, 
impurities, ions on wrong lattice sites, etc.). 
The systems are too large to make first-principle 
calculations that consider these effects, although
first attemps already yield impressive
fundamental predictions \cite{Phillpot04Coup}.

The simplest periodic structure, that is ideally 
suited to compare theoretical and experimental 
data, is an elementary grating. Such periodically 
poled crystals are also of relevance for a variety of 
applications: 
Frequency conversion \cite{Franken61Gene}, parametric 
oscillation \cite{Dunn99Para}, the generation of terahertz 
radiation \cite{Lee00Gene} as well as high-speed electro-optical 
switching \cite{Yamada00Elec}. Conventional periodic poling 
is achieved by applying high electrical fields with 
structured electrodes \cite{Yamada93Firs}. However, 
quality and size of domain structures fabricated 
by this method seem to be limited. The reasons are not 
clear because of insufficient theoretical modelling 

One possibility to describe ferroelectric domain 
reversal is the theory of Avrami, Kolmogorov, 
Johnson, and Mehl on first-order phase transitions 
adapted by Sekimoto to a form usable for our purposes 
\cite{Ishibashi71Note, Sekimoto91Evol}. This theory 
relys on a purely stochastical description. According to 
it, domains nucleate and grow independently from 
each other, and within its framework the 
probability for domain inversion can be calculated. 
Usually the domain switching is studied by obtaining 
the time dependence of the inverted volume fraction 
during the poling process. It is an open and demanding 
question whether this theory can be applied to domain 
patterning as well. 

In the present work electrical fixing 
\cite{Micheron73Fiel,Qiao93Elec,Cudney93Phot,Eggert04Elec}
is used to compare stochastical modelling of
domain patterning to experimental data: 
Space-charge fields are created in a crystal 
via the photorefractive effect 
\cite{Buse97Ligh1}, and additionally an external 
electrical field is applied. According to the 
stochastical model a spatially modulated 
electrical field leads to a modulated 
probability for domain inversion. The 
dependence of the domain grating quality on 
the underlying probability density can be calculated
by adapting the theory of Avrami, Kolmogorov, 
Johnson, and Mehl to the case of a spatially 
sinuosoidally modulated probability for domain 
nucleation. Sekimoto 
\cite{Sekimoto91Evol} gives the following general 
formula for the probability $w$ that a point 
$\vec{r}$ has an inverted spontaneous polarization:
\begin{equation*}
w(\vec{r},t) = 1-\exp\left(-\int_0^t \mathrm{d}t' 
\int_{V} \mathrm{d}^3r' I(\vec{r'},t')\left[1-D(\vec{r},t;\vec{r'},t')\right] \right)\,.
\end{equation*} 
Here $I(\vec{r'},t')$ denotes the probability density 
for domain nucleation at position $\vec{r'}$ and 
time $t'$ and $V$ denotes the whole crystal volume.
The function $D(\vec{r},t;\vec{r'},t')$ is zero if 
such a domain contains $\vec{r}$ at time $t$ and is 
one otherwise. The time dependence in $D$ and $I$ is 
introduced to model the temporal evolution as domain 
nucleation probability and domain sizes may vary with 
time. As we are not interested in the temporal evolution 
here, we replace the time dependence by the dependence 
on domain length $l\,.$ Thus it follows:
\begin{equation}\label{mastereq}
w(\vec{r}) = 1-\exp\left(-\int_0^{\infty} \mathrm{d}l' 
\int_V \mathrm{d}^3r' I(\vec{r'},l')\left[1-D(\vec{r},\vec{r'},l')\right] \right)\,.
\end{equation} 
Here $I(\vec{r'},l')$ is the probability that a 
domain located at $\vec{r'}$ with length $l'$ can 
be found.  

We now assume that due to a periodic 
space-charge grating the probability
for domain nucleation is periodically 
modulated. Thus we write: 
\begin{equation}\label{Iform}
I(\vec{r'},l')=[\alpha+\beta\sin(K r'_z)]n_l(l')\,.
\end{equation} 
Here $n_l(l')$ describes the size distribution
of domains. 
So $\beta/\alpha$ is a measure for the strength 
of the influence of the space-charge grating 
on domain nucleation. Here $K=2\pi/\Lambda$ is the 
spatial frequency of the space-charge grating. The domain 
shape is assumed to be cylindrical with length $2l$
and area $A(l)$ and to be centered around $\vec{r'}$ 
which is described by the function $D\,.$ Instead of 
a modulated probability for domain nucleation, a 
modulated domain-wall growth can be assumed by 
adapting $I(\vec{r'},l')\,.$ However, within 
the framework of our methods both assumptions lead
to very similar predicitions, hence the latter
is not investigated within this article. 
Inserting equation (\ref{Iform}) into 
equation (\ref{mastereq}) and using the specific 
domain form, we get: 
\begin{equation*}
\begin{split}
w(\vec{r}) = 1-& \exp\Big(-2\alpha \int_0^{\infty} \mathrm{d}l\, 
n_l(l) A(l) l \Big)\\
&\times \exp\Big( -2\sin(Kr_z)\frac{\beta}{K} 
\int_0^{\infty} \mathrm{d}l\, n_l(l) A(l) \sin(Kl) \Big)\,.
\end{split}
\end{equation*} 
We define $\gamma=2\alpha \int_0^{\infty} \mathrm{d}l\, n_l(l) A(l) l$
and $\eta=2(\beta/K) \int_0^{\infty} \mathrm{d}l\, n_l(l) A(l) \sin(Kl)$.
The former is a measure for the volume of inverted spontaneous 
polarization whereas $\kappa=\eta/\gamma$ measures the modulation 
degree of the domain grating ($0\leq\kappa\leq 1$). 
Expanding the second exponential function 
in the last expression for $w$ up to the 
first order we get as an approximate 
expression for $w$:
\begin{equation*}
\begin{split}
w(\vec{r}) = 1-\left[ 1-\gamma\kappa\sin\left(Kr_z\right) \right]\exp\left(-\gamma \right)\,.
\end{split}
\end{equation*} 
The area fraction $q$ with inverted spontaneous polarization 
(degree of poling) can be found by taking the spatial average 
of $w(\vec{r})$:
\begin{equation*}
\begin{split}
q = \frac{1}{\Lambda} \int_0^{\Lambda} \mathrm{d}z\, q(z) = 1-\exp\left(-\gamma \right)\,.
\end{split}
\end{equation*}
The degree of modulation of the domain grating $\Delta q$ can be calculated 
as the first Fourier coefficient for $K$:
\begin{equation}\label{finalresult}
\begin{split}
\Delta q = \gamma \kappa \exp(-\gamma) =-\kappa(1-q)\ln(1-q)\,.
\end{split}
\end{equation} 

Equation \ref{finalresult} expresses the domain grating modulation solely as a 
function of the degree of poling. For no or complete reversal of the 
spontaneous polarization no grating at all can be found. A maximum 
is found in between. Numerically evaluating equation 
\ref{finalresult} for several assumptions yields that 
for the height of the curve 
$\max\Delta q\approx 0.4 \kappa$ roughly holds. 

To test these predictions experimentally, electrical 
fixing experiments are performed. Lithium 
niobate crystals doped with $0.05\,\mathrm{mol\%}$ 
iron from Deltronic Crystal Industries are used. 
To reduce the coercive field a vapor transport 
equilibration treatment is applied \cite{Jundt90Opti}. 
The crystals are put into a special mount which 
is placed in a glass bin filled 
with silicon oil to prevent electrical breakdowns.  

The optical setup consists of a detuned Mach-Zehnder interferometer, 
which allows to illuminate the crystal with a periodic interference 
pattern with grating periods of $5$ to $100 \,\mathrm{\mu m}.$ 
Light of the wavelength $488\,\mathrm{nm}$ from an Argon-ion laser 
is used. The illumination creates a space-charge field 
which induces an index-of-refraction grating via the electro-optic
effect. This photorefractive volume grating is detected by 
Bragg-diffraction with light of the wavelength $633\,\mathrm{nm}$ 
from a HeNe-laser. From the amount of diffracted light the amplitude 
of the modulation of the index-of-refraction and hence of the 
space-charge field can be calculated \cite{Kogelnik69Coup}. A 
detailed description of the setup can be 
found in reference \cite{Eggert05Opti}. 

The experiments are performed as follows: 
A space-charge field is written by illuminating the 
crystal with an interference grating. Either during or 
after the recording a voltage is applied to the crystal 
along the z-axis. The polarity of the voltage is 
chosen such that the electrical field supports domain 
inversion. The strength of the space-charge 
field is monitored during the whole experiment. When the 
space-charge field reaches steady state, the 
illumination is stopped and the external electrical field 
is shut off. After homogeneous illumination of the 
crystal no diffracted light is detected. Next the 
crystal is repoled to its single-domain state by 
applying an electrical field without any 
illumination. The degree of poling is determined 
by integrating the current during the 
poling process. This value is normalized by twice
the spontaneous polarization $P_{\mathrm{S}}$.

\begin{figure}
\includegraphics[width=\columnwidth]{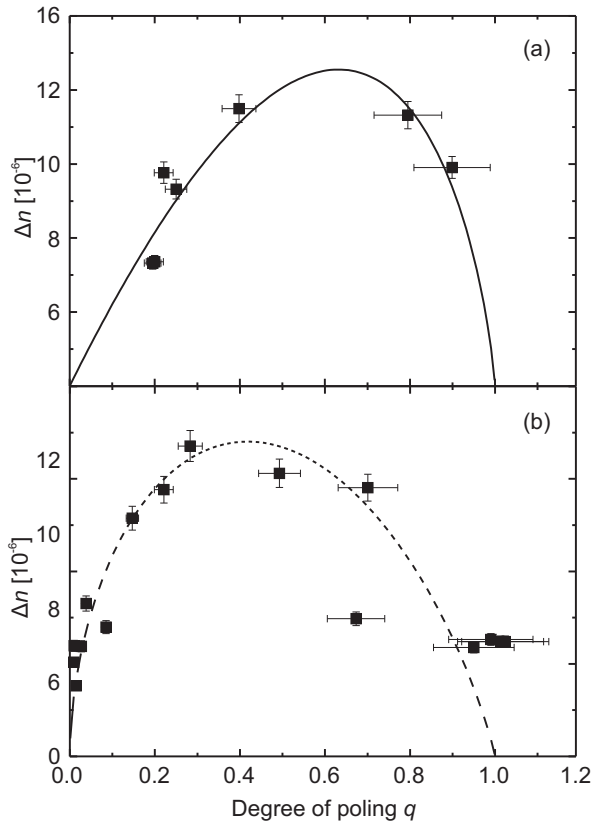}
\caption{\label{fig:DNFIELD}
The index-of-refraction modulation $\Delta n$ 
is plotted versus the degree of poling $q.$ 
The period $\Lambda$ of the fixed grating is 
$\Lambda=15.7\,\mathrm{\mu m}.$ In 
Fig.\,\ref{fig:DNFIELD}\,(a) the degree of 
poling is varied by applying an external 
electrical field for different times. The 
solid line is a plot of Eq.\,(\ref{finalresult}) 
where $\kappa$ is a fit parameter. In 
Fig.\,\ref{fig:DNFIELD}\,(b) different external 
fields are applied for a constant time.
The dashed line is a guide to the eye.}
\end{figure}

The revealed index-of-refraction grating provides information 
about the achieved degree of modulation  of the domain walls.
It is studied for different strengths and application 
times of the external field during the recording process. 
In Fig.\,\ref{fig:DNFIELD} the revealed space-charge 
grating strength is plotted against the degree of poling.
In Fig.\,\ref{fig:DNFIELD}\,(a) a space-charge field is 
written in the crystal before the external field 
is applied. The experimental parameter, which is varied, 
is the time this field is applied. In 
Fig.\,\ref{fig:DNFIELD}\,(b) the space charge grating is 
written in the presence of an externally applied 
field. Here the recording time is fixed but in order to 
change the degree of poling the external field 
is varied. 

As can be seen, the strength of the revealed 
field becomes zero for $q\rightarrow 0$ as 
well as for $q\rightarrow 1$ and reaches a 
maximum in between. 
In Fig\,\ref{fig:DNFIELD}\,(a) the dependence 
of the strength of the reappearing space-charge 
field is shifted to higher $q$-values compared 
to the results obtained 
with varying fields in Fig.\,\ref{fig:DNFIELD}\,(b).

Comparing theoretical predicitions with experimental 
data for the strength of the revealed field under 
different fixing conditions, qualitative
agreement can be found. The dependence on the 
degree of poling $q$ with different writing times 
as in Fig.\,\ref{fig:DNFIELD}\,(a) is 
very close to a $-(1-x)\mathrm{ln}(1-x)$ dependence
as predicted by Eq.\,(\ref{finalresult}). The shift 
to lower $q$-values, when the field is varied, is 
understandable: Different external fields do not 
only influence $q$ but also change
the relative influence of the space-charge field 
and thus $\beta/\alpha$.

\begin{figure}
\includegraphics[width=\columnwidth]{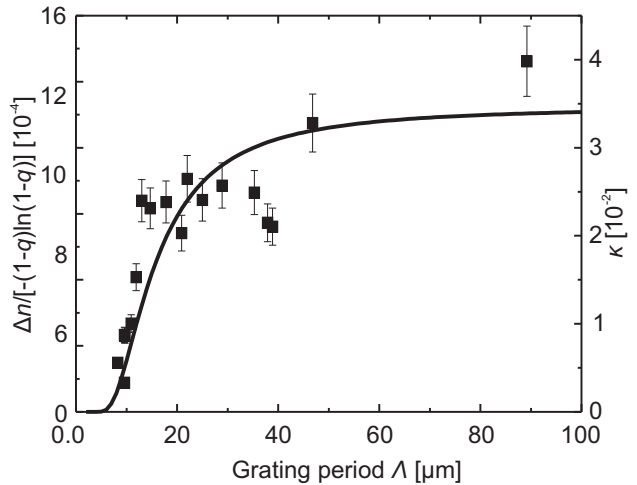}
\caption{\label{fig:DNLAMBDA}
The index-of-refraction modulation $\Delta n$ is 
plotted versus the period length $\Lambda$ of the 
fixed grating. To correct for different 
degrees of poling, $\Delta n$ was divided 
by $-(1-q)\ln(1-q)$ (e.g. Eq.\,(\ref{finalresult})).  
The curve is the product of a fit parameter 
$A$ and the caluclated $\kappa$, which 
determines the degree of modulation $\Delta q$.
A Gaussian distribution is assumed for the 
distribution of domain lengths.}
\end{figure}
Experimental data for different grating 
periods is shown in Fig.\,\ref{fig:DNLAMBDA}. For 
period lengths smaller than $15\,\mathrm{\mu m}$
the strength of the reappearing space-charge field 
quickly drops to zero. However, for large period 
lengths exceeding $40\,\mathrm{\mu m}$ saturation 
is obvious. 

Fig.\,\ref{fig:DNLAMBDA} not only shows the measured 
strengths of the revealed index-of-refraction grating 
for different period lengths $\Lambda$, but also a
calculation of the degree of modulation. Here assumptions 
on the domain size distribution have to be made: 
A Gaussian distribution centered at $0$ with a 
width of $5\,\mathrm{\mu m}$ is assumed. The height of 
the plotted curve is used as a fit parameter. The basic 
message is that if a broad domain size distribution
is assumed, as it is found in real crystals by the 
PFM measurements, the model predicts that no 
revealed grating should be found for period lengths 
smaller than the length of most of the domains. 
For higher period lengths the grating quality 
increases. These predictions are in  
agreement with measured data. 

\begin{figure}
\includegraphics[width=\columnwidth]{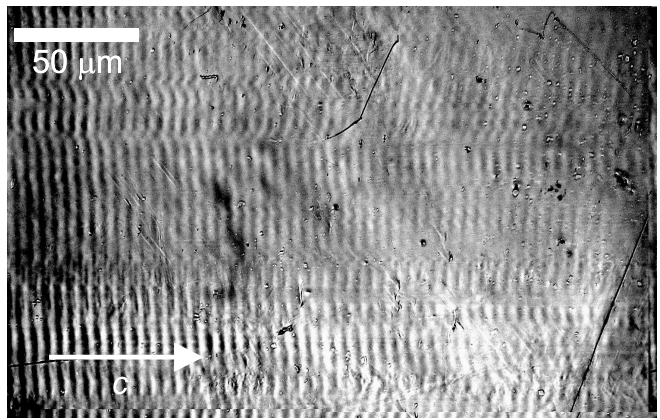}
\caption{\label{fig:DICIMG}
A revealed grating is imaged with the differential 
interference contrast method. The period length 
$\Lambda$ of the electrically fixed grating is 
$9.6\,\mathrm{\mu m}.$ The z-axis is aligned
horizontally.}
\end{figure}
To check whether a space-charge grating is revealed 
all over the crystal or only in some parts of it, 
the index grating induced by the space-charge field 
is imaged with differential interference contrast 
microscopy (DIC) after repoling the crystal. This method 
allows to detect very small gradients of the 
index-of-refraction in the crystals. We use this 
as a method to observe restored space-charge fields 
and to determine qualitatively the shape of fixed 
domain gratings. This is possible because after 
repoling of the crystal uncompensated charges 
appear at the former domain walls. These charges 
cause electro-optic index changes which are a 
replica of the original domain pattern \cite{Qiao93Elec}. 
Figure\,\ref{fig:DICIMG} 
shows the result. As it can be seen, the grating
reappears all over the crystal. 

\begin{figure}
\includegraphics[width=\columnwidth]{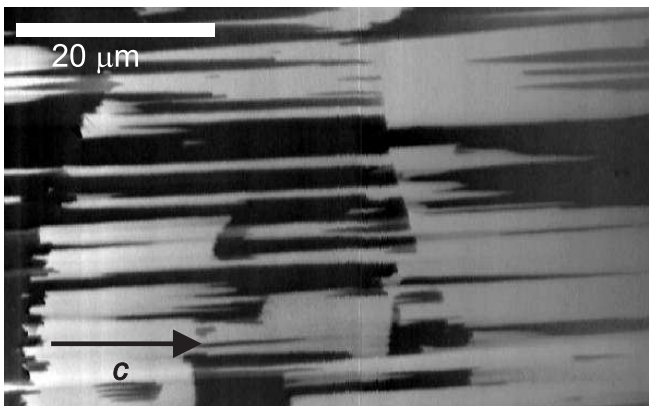}% Here is how to import EPS art
\caption{\label{fig:AFM2}
A fixed domain grating with a period length 
$\Lambda$ of $11.7\,\mathrm{\mu m}$ is imaged by
piezoelectric force microscopy. The two 
grey values represent the two directions of the 
spontaneous polarization. The z-axis is 
aligned horizontally.}
\end{figure}
Piezoelectric force microspcopy (PFM) is used 
to directly image the ferroelectric domain 
structure after electrically fixing a 
space-charge grating. 
The PFM is a scanning force microscope operated in contact
mode with an additional alternating voltage applied to the tip
\cite{Alexe}. In piezoelectric samples this voltage causes thickness
changes and therefore vibrations of the surface which lead to
oscillations of the cantilever that can be read out with a lock-in
amplifier. In the present work we imaged the non-polar faces of
LiNbO$_3$.

The faces of the crystal are polished down 
stepwise by some hundred micrometers to ensure
that volume domains are visualized. 
Figure\,\ref{fig:AFM2} shows that 
ferroelectric volume domains can be found after 
electrical fixing of holograms in lithium niobate crystals.  
However, a periodic structure is not immediately 
evident. In fact, the length of some areas with 
inverted spontaneous polarization are bigger
than one period of the fixed grating. 

However, from the PFM images we conclude that 
electrical fixing indeed leads to ferroelectric volume 
domains with inverted spontaneous polarization
in head-to-tail configuration, but the periodicity 
is not evident locally. The DIC images suggest 
that averaged over the whole crystal thickness 
a well defined grating is present. Both 
observations are
consistent with the predictions of the model. 
The degree of modulation is too small to be 
obvious from direct domain visualization,
but an average periodicity is visible which 
can be detected by the help of light diffraction
at the revealed space charge grating. 

After revealing, local fields are created according 
to the former domain structure. Typical values of 
such fields $E_{\mathrm{loc}}$ presumably are 
determined by material parameters, as the spontaneous 
polarization and field-limiting effects like 
the electric-breakdown field. The $E_{\mathrm{loc}}$ 
form a corresponding index-of-refraction grating via 
the electro-optic effect which is detected by 
diffraction of a light beam. The effective, averaged
field $E_{\mathrm{eff}}$, which is measured by 
this method, is given by the modulation degree 
multiplied by the local electric field 
$E_{\mathrm{loc}}$. Thus theoretical 
predictions for $\Delta q$ link a material 
dependent factor $E_{\mathrm{loc}}$ with the 
experimental value $E_{\mathrm{eff}}$, i.e., these
values are proportional to each other.

Alltogether it is found that the predictions of the 
model are in agreement with all available 
experimental data, indicating strong support for 
the model presented herein. None of the assumptions 
of the theory is specific to the material lithium 
niobate. So the stochastic description offers a 
general way to analyze and discuss experimental 
dependences in all electrical fixing experiments 
which have been performed in a large variety of 
ferroelectric materials. For Example in 
Ref.\,\cite{Cudney06Elec}
the same dependence of revealed grating strengths 
on the degree of poling and the same dependence on 
the grating period $\Lambda$ is found for electrical 
fixing in barium titanate crystals.

Furthermore, the stochastical description enables
quantitative predictions. Provided better knowledge 
of local fields after revealing is gained, the factor 
$\beta/\alpha$, i.e.\ the influence of local 
fields on the poling dynamics, can be analyzed. 
This is a point which is hard to access by 
alternative means. As the influence of defects 
presumably can be described by local fields, 
deeper understanding of ferroelectric domain 
switching becomes possible. 

Financial support by the DFG (BU\,913/11), by the Deutsche 
Telekom AG, and by the Deutsche Telekom Stiftung is gratefully 
acknowledged. 

\newcommand{\noopsort}[1]{} \newcommand{\printfirst}[2]{#1}
  \newcommand{\singleletter}[1]{#1} \newcommand{\switchargs}[2]{#2#1}

\end{document}